\documentclass[aps,prl,twocolumn,amsmath,showpacs,superscriptaddress]{revtex4}
\usepackage{graphicx}
\usepackage{bm}
\usepackage{amssymb}

\begin{document}

\title{Angular Momentum of Phonons and Einstein-de Haas Effect}

\author{Lifa~Zhang}
\affiliation{Department of Physics, The University of Texas at Austin, Austin, Texas 78712, USA}

\author{Qian~Niu}
\affiliation{Department of Physics, The University of Texas at Austin, Austin, Texas 78712, USA}
 \affiliation{International Center for Quantum Materials, Peking University, Beijing 100871, China}

\date{27 Jan 2014}

\begin{abstract}
We study angular momentum of phonons in a magnetic crystal. In the presence
of a spin-phonon interaction, we obtain a nonzero angular momentum of phonons,
which is an odd function of magnetization. At zero temperature,  phonon has a
zero-point angular momentum besides a zero-point energy. With increasing
temperature, the total phonon angular momentum diminishes and approaches to zero
in the classical limit. The nonzero phonon angular momentum can have a
significant impact on the Einstein-de Haas effect. To obtain the change of
angular momentum of electrons, the change of phonon angular momentum needs to be
subtracted from the opposite change of lattice angular momentum. Furthermore, the finding of phonon angular momentum gives a potential method to study the spin-phonon interaction. Possible experiments on phonon angular momentum are also discussed.
\end{abstract}

\maketitle
The Einstein-de Haas effect \cite{einde,frenkel79}, a phenomenon of mechanical
rotation induced by a magnetization change, was originally designed to prove the
existence of Ampere's molecular currents; but subsequent experiments \cite{stewart18} showed that the magnetic moment of an atom is dominated by spin while contribution from orbital motion to the magnetic moment is almost absent. The Einstein-de Haas experiment together with the Barnett experiment \cite{barn15,barn35} (a change of magnetization resulting from a mechanical rotation)  has provided an effective method of measuring the gyromagnetic ratio for various materials \cite{davi53,kitt49,reck69}. The accuracy
of gyromagnetic ratio is crucial to determining of orbital and spin contribution
in total magnetization \cite{stoh95,ulme04,lidb09,boeg10,pere11, niem12,dzya11}.

Due to conservation of total angular momentum of the whole system in the Einstein-de Haas effect, the change of angular momentum of electrons (including both spin and orbital parts) has taken to be equal in magnitude but opposite in sign to the change of lattice angular momentum, which corresponds to mechanical rotation. However, the mechanical rotation only
reflects angular momentum of the rigid-body lattice where atoms are assumed in
the corresponding equilibrium positions; while phonons, which come from atomic
vibrations around equilibrium positions, are assumed to have no macroscopic angular
momentum. Recently, a remarkable phenomenon of phonon Hall effect was observed in a
paramagnetic insulator \cite{phee1,phee2}, which is indeed a surprise since phonons
as neutral quasi-particles cannot directly couple to magnetic field via Lorentz
force. The following theoretical studies \cite{phet1,phet2} showed that
through Raman spin-phonon interaction the magnetic field can have an effective
force to distort phonon transport, and thus drive a
circulating heat flow \cite{qin11}. Therefore a natural question arises:
can such circulating phonons have nontrivial angular momentum and emergent
macroscopic effects?

In this Letter, we study angular momentum of phonons in a magnetic crystal in
a microscopic picture. It is found that the Raman spin-phonon interaction induces a
nonzero phonon angular momentum, which is an odd function of magnetization. In
addition to a zero-point energy, phonon has a zero-point angular momentum
at zero temperature.   Such zero-point phonon angular momentum is offset by that of
excited phonon modes such that the total angular momentum of phonons vanishes in
the classical limit. Phonon angular momentum can not be ignored in total angular momentum especially in magnetic materials with large magnetization and spin-phonon interaction. Revisiting the Einstein-de Haas effect we find that phonon
angular momentum needs to be subtracted in calculating angular momentum of
electrons. With this correction, spin and orbital angular momentum can be
precisely determined. Besides the Einstein-de Haas effect, nontrivial phonon angular momentum can be applied to the study of spin-phonon interaction, thermal Hall effect, and other topics related to phonons.

{\it Angular momentum of phonons} -- The lattice angular momentum related to
mechanical rotation only reflects the rigid-body motion of the lattice. However, angular momentum of phonons has never been considered. In a microscopic picture, we can define angular momentum of phonons as
\begin{equation}\label{eq_angph0}
\bm J^{\rm ph}=\sum_{l\alpha} \bm u_{l \alpha} \times \dot{\bm u}_{l \alpha}.
\end{equation}
Here $\bm u_{l \alpha}$ is a displacement vector of the $\alpha$-th atom in the
$l$-th unit cell, multiplied by square root of mass.
Along $z$ direction, $J^{\rm ph}_z=\sum_{l\alpha} (u_{l \alpha}^x \dot{u}_{l
\alpha}^y-u_{l \alpha}^y \dot{u}_{l \alpha}^x)$.  One can present the displacement
in the second quantization form as $u_l = \sum_{k} \epsilon_k e^{i ( {\bf R}_l
\cdot {\bm k}-\omega_k t)} \sqrt{\frac{\hbar}{2\omega_k N}}\; a_k + {\rm h.c.}$,
with $k=(\bm k, \sigma)$ specifying a wave vector $\bm k$ and a branch $\sigma$,
where  $\epsilon_k$ is a displacement polarization vector. Then the phonon angular
momentum can be written as \cite{supp}
{\small
\begin{eqnarray}
J_z^{\rm ph} & =&\frac{\hbar}{2}
\sum_{k,k'} \epsilon_k^\dagger M \epsilon_{k'}  \left(
\sqrt{\frac{\omega_k}{\omega_{k'}}}+
\sqrt{\frac{\omega_{k'}}{\omega_k}}\right)a_k^\dagger a_{k'}
\delta_{{\bm k},{\bm k}'} e^{i(\omega _k  - \omega _{k'} )t} \nonumber \\ &
&+\frac{\hbar}{2}\sum_k\epsilon_k^\dagger M \epsilon_{k}.
\end{eqnarray}}
Here $M=\left( \begin{smallmatrix} 0 & -i \\ i & 0 \end{smallmatrix}
\right)\otimes I_{n\times n} $, and $n$ is the number of atoms in one unit cell.
In equilibrium, the angular momentum of phonons reduces to   \cite{supp}:
\begin{equation}\label{eq_angph}
J^{\rm ph}_z=\sum_{\sigma,\bm k} l_{\bm k, \sigma}^z\, [f(\omega_{\bm k,
\sigma})+\frac{1}{2}],\; \;l_{\bm k, \sigma}^z=(\epsilon^\dagger_{\bm k,
\sigma}\,M\,\epsilon_{\bm k, \sigma})\hbar,
\end{equation}
where {\small $f(\omega_k)=\frac{1}{e^{\hbar\omega_k/k_B T}-1}$ } is the
Bose-Einstein distribution. In Eq.(\ref{eq_angph}), we do summation over all wave
vector points and all phonon branches ($\omega\geq0$). Here $l_{\bm k, \sigma}^z$
is the phonon angular momentum of branch $\sigma$ at wave vector $\bm k$,  which is
 real and proportional to $\hbar$. At zero temperature,  the total phonon
angular momentum is
$J^{\rm ph}_z(T=0)=\sum_{\sigma,\bm k} \frac{1}{2}\,l_{\bm k, \sigma}^z $, which
means that each mode of $(\bm k, \sigma)$ has a zero-point angular momentum {\small
$\frac{1}{2} l_{\bm k, \sigma}^z= \frac{\hbar}{2} (\epsilon^\dagger_{\bm k,
\sigma}\,M\,\epsilon_{\bm k, \sigma}) $ } besides a zero-point energy of
$\hbar\omega_{\bm k, \sigma}/2$.

For an ionic crystal lattice in a uniform external magnetic field, the Hamiltonian
reads in a compact form \cite{phet1,phet2,holz72,wang09}:
\begin{equation}\label{eq_ham}
H  = \frac{1}{2} (p-{ \tilde A }
u)^T (p-{ \tilde A }u) + \frac{1}{2} u^T K u,
\end{equation}
where $u$ is a column vector of
displacements from lattice equilibrium positions, multiplied by square root of mass; $p$ is
a conjugate momentum vector, and $K$ is a force constant matrix.  The cross term
$u^T{ \tilde A }p$ can be interpreted as a Raman spin-phonon interaction
\cite{ray67,iose95}.
The superscript {\small $T$} stands for the matrix transpose. ${ \tilde A }$,
an antisymmetric real matrix \cite{qin12}, has a dimension of $N\,d\times N\,d$ where $N$ is the number of total sites and $d$ is the dimension of lattice vibrations;  in a proper approximation it can be block diagonal with elements  $ \Lambda_\alpha=
\left(\begin{smallmatrix} 0 & \lambda_\alpha  \\
-\lambda_\alpha & 0  \\
\end{smallmatrix}\right)
$ with respect to the $\alpha$-th ionic site,  where we only consider two-dimensional ($x$ and $y$
directions) motion of the lattice ($d=2$). Here $\lambda_\alpha$ has a dimension of frequency, and is proportional to the
spin-phonon interaction and magnetization,  which is assumed to be proportional to magnetic field for a paramagnetic material. The magnetic field is applied
along $z$ direction. The polarization vector $\epsilon$ satisfies
$\bigl[ (-i\omega + A)^2 + D\bigr] \epsilon = 0,$ where $D({\bf k}) = -
A^2+\sum_{l'} K_{ll'} e^{i({\bf R}_{l'} - {\bf R}_{l})\cdot {\bf k}}$ is the
dynamic matrix and $A$ is block diagonal with the element of $\Lambda_\alpha$, and has a dimension of  $2n\times 2n$ where $n$ is the number of sites
per unit cell.

In absence of spin-phonon interaction, the system reduces to a trivial phonon
system $
H  = \frac{1}{2} p^Tp + \frac{1}{2} u^T K u$. Solving the simple eigenvalue problem
as $D(\bm k)\epsilon_{\bm k, \sigma}=\omega_{\bm k, \sigma}^2 \epsilon_{\bm k,
\sigma}$ with $D^T({\bf k})=D^*({\bf k})=D(-{\bf k})$, one can have $\omega_{-\bm
k, \sigma}=\omega_{\bm k, \sigma}, \epsilon_{-\bm k, \sigma}=\epsilon_{\bm k,
\sigma}^*$, then we obtain $ l_{-\bm k, \sigma}^z= -l_{\bm k, \sigma}^z$ and
$J^{\rm ph}_z=0$ \cite{supp}. Thus for a phonon system without a spin-phonon
interaction, the total angular momentum of phonons is zero.
\begin{figure}[t]
\includegraphics[width=3.20 in,  angle=0]{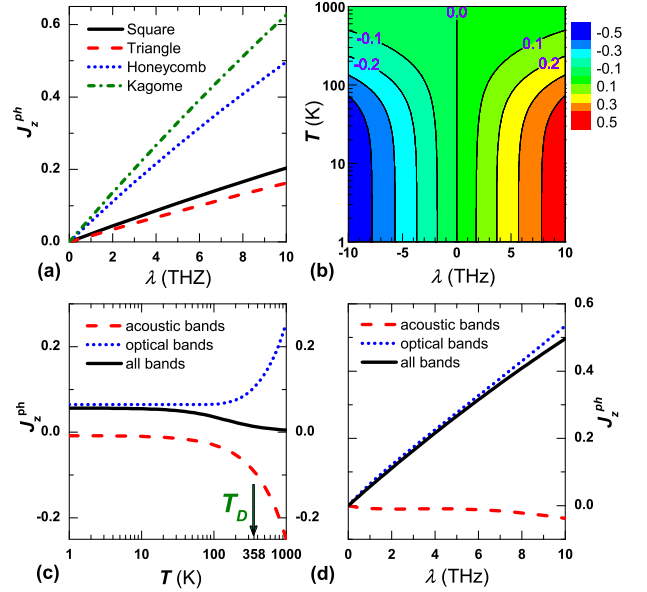}
\caption{ \label{fig1} (Color online) (a) The phonon angular momentum $J_z^{\rm ph}$ of one unit cell as a function of $\lambda$ at temperature $T=0$ K for different lattice symmetries. (b) The contour plot of the phonon angular momentum $J_z^{\rm ph}$ of one unit cell  as a function of $\lambda$ and temperature $T$. (c) The phonon angular momentum $J_z^{\rm ph}$ of one unit cell from different phonon bands as a function of temperature $T$ at $\lambda=$ 1 THz, where the arrow denotes the Debye temperature of the model ($T_D=358$ K).  (d) The phonon angular momentum $J_z^{\rm ph}$ of one unit cell from different phonon bands as a function of $\lambda$ at $T=0$ K. The phonon angular momenta in (b)-(d) are calculated for a honeycomb lattice. All the phonon angular momenta are in the unit of $\hbar$.}
\end{figure}

For a phonon system with a spin-phonon interaction, $\epsilon_{-\bm k,
\sigma}=\epsilon_{\bm k, -\sigma}^*\neq \epsilon_{\bm k, \sigma}^*$, and then $
l_{-\bm k, \sigma}^z \neq-l_{\bm k, \sigma}^z$, thus one can get a nonzero phonon
angular momentum, which is shown in Fig.~\ref{fig1}. We calculate phonon angular
momentum for lattices with the following parameters: the longitudinal spring constant is $K_L = 0.144\,$eV/(u\AA$^2$) and the transverse one is $K_T=K_L/4$; the unit cell lattice vectors are $(a,0),\,(0,a)$ for a square lattice and $(a,0), \,(a/2, a\sqrt{3}/2)$ for other lattices with $a=1\,$\AA. We take $\lambda_\alpha=\lambda$ for the model calculation \cite{comment2}.  Figure~\ref{fig1}(a) shows that honeycomb and kagome lattices have larger phonon angular momenta than those of triangle and square lattices, which means that lattices with more sites per unit cell can have a larger phonon angular momentum. We can understand this trend by observing that optical bands are more important in contributing to the phonon angular momentum than the acoustic ones. In Fig.~\ref{fig1}(c) and (d) we plot the phonon angular momentum contributing from different bands in a honeycomb lattice. It is shown that the phonon angular momentum from acoustic bands almost vanishes at low temperatures (see Fig.~\ref{fig1}(c)) and if $\lambda$ is not large (see Fig.~\ref{fig1}(d)), thus the optical bands dominate the contribution to the total phonon angular momentum. With more sites per unit cell more optical bands present, thus phonon angular momentum will be larger.

By using the relations {\small $\epsilon_{- \bm k, \sigma}^* (-A) = \epsilon_{\bm
k, \sigma} (A)$, $\omega_{-\bm k, \sigma}(-A) = \omega_{\bm k, \sigma} (A)$,
$M^T=-M$}, we can obtain $J^{\rm ph}_z (-\lambda)=-J^{\rm ph}_z(\lambda)$. Since
$\lambda$ is proportional to magnetization, the total angular momentum of
phonon will change sign when magnetization changes sign.  As shown in
Fig.~\ref{fig1}(a), (b) and (d), the total angular momentum of phonons per unit cell  increases as $\lambda$ increases; but the increase rate will decrease.

{\it Angular Momentum in the Classical Limit} -- At the high temperature limit,
from Eq.~(\ref{eq_angph}) we have \cite{supp}:
\begin{equation}\label{eq_angtinf}
J^{\rm ph}_z(T\rightarrow\infty)=\sum_{\sigma>0,\bm k} [ (\frac{k_B
T}{\hbar\omega_{\bm k, \sigma}}+\frac{\hbar\omega_{\bm k, \sigma}}{12 k_B T}
)l_{\bm k, \sigma}^z ].
\end{equation}
It seems that the phonon angular momentum would be linear with temperature at the
high temperature limit. However, the first term vanishes due to the fact of  {\small $ \sum_{\sigma>0,\bm k} \frac{\epsilon^\dagger_{\bm k,
\sigma}\,M\,\epsilon_{\bm k, \sigma}}{\omega_{\bm k, \sigma}}=0$} \cite{supp}. Therefore at a high temperature total phonon angular momentum is proportional to $1/T$ and tends to zero as
\begin{equation}
J^{\rm ph}_z(T\rightarrow\infty) =\sum_{\sigma>0,\bm k}  \frac{\hbar\omega_{\bm k,
\sigma}}{12 k_B T}l_{\bm k, \sigma}^z \rightarrow 0.
\end{equation}
The phonon angular momentum per unit cell changing with temperature is shown in
Fig.~\ref{fig1}(b) and (c). Whatever a magnetic field is applied,  the phonon angular momentum per unit cell decreases with increasing temperature and tends to zero at the high temperature limit ($T \gg T_D$). With increasing temperature more modes are exited, the angular momentum of which has the direction opposite to
that of zero-point angular momentum; at the high temperature limit, the phonon
angular momentum of all the excited modes exactly cancels out the zero-point
angular momentum ({\small $\sum l_{\bm k, \sigma}^z\, f(\omega_{\bm k,
\sigma},T\rightarrow\infty)=- \sum \frac{1}{2}\,l_{\bm k, \sigma}^z $ }).
We can understand the absent phonon angular momentum in the classical limit as
follows. At high temperatures, classical statistical mechanics is applicable to
calculate phonon angular momentum.   Summation over quantum states becomes a
phase-space integral with respect to $p$ and $u$.  One can do a change of variable
to make the kinetic energy in the Hamiltonian Eq.~(\ref{eq_ham}) into a usual form
$p^2/2$, thus removing the effect of ${ \tilde A }u$; for such a pure harmonic
system, the angular momentum of phonons is zero as discussed above. Furthermore,
the Bohr–van Leeuwen theorem states that in classical mechanics the thermal average
of the magnetization is always zero \cite{bvl}, which also makes the angular
momentum of phonons vanish at the classical limit. Therefore, the phonon angular
momentum is meaningful only in low-temperature quantum systems.

{\it Revisit the Einstein-de Haas Effect} -- The Einstein-de Haas effect
\cite{einde} showed a mechanical rotation of a freely suspended body
caused by the change in its magnetization.  In their experiment \cite{einde},
Einstein and de Haas employed a resonance method in which the magnetic field was
periodic and tuned to be the natural frequency of the rod and its suspension, which
provided measurements for the ratio between the change in magnetization and the one
in total angular momentum. Traditionally the total angular momentum is assumed as
$\bm J^{\rm tot}=\bm J^{\rm lat}+\bm J^{\rm spin}+\bm J^{\rm orb} $, thus due to
conservation of angular momentum, one obtains $\Delta\bm J^{\rm lat}=-(\Delta\bm
J^{\rm spin}+\Delta\bm J^{\rm orb})$ which is determined by the mechanical rotation
of the sample \cite{kitt49}. However, from a microscopic point of view, the
angular momentum of all atoms in the sample can be written as
\begin{equation}
\bm J^{\rm atom}=\sum_{l\alpha} (\bm R_{l \alpha} +\bm u_{l \alpha} )\times(
\dot{\bm R}_{l \alpha} +\dot{\bm u}_{l \alpha}),
\end{equation}
where $\bm R_{l \alpha}$ is the equilibrium position of the $\alpha$-th atom in the
$l$-th unit cell, multiplied by square root of its mass. The angular
momentum of lattice is
 \begin{equation}\label{eq_angl}
 \bm J^{\rm lat}=\sum_{l\alpha} \bm R_{l \alpha} \times \dot{\bm R}_{l \alpha},
\end{equation}
which really reflects the mechanical rotation of rigid-body motion of the sample.
In equilibrium, the cross terms related with $\bm u$ or $\dot{\bm u}$ are zero, then $\bm J^{\rm atom}=\bm J^{\rm lat}+\bm J^{\rm ph}$. Thus the total angular momentum
should be
\begin{equation}\label{eq_angt}
 \bm J^{\rm tot}=\bm J^{\rm lat}+\bm J^{\rm ph}+\bm J^{\rm spin}+\bm J^{\rm orb}.
\end{equation}
The global conservation of angular momentum does not explain how the angular
momentum is actually transferred from individual electrons or atoms to the whole
rigid body; the Raman type spin-phonon interaction can be ubiquitous and
plays an essential role. According to the discussion in the above section, we know
that in the presence of the spin-phonon interaction, the phonon band structure is
nontrivial and gives nonzero angular momentum $\bm J^{\rm ph}$. Based on
conservation of total angular momentum we obtain
\begin{equation}\label{eq_deltj}
\Delta\bm J^{\rm spin}+\Delta\bm J^{\rm orb}=-\Delta\bm J^{\rm lat}-\Delta\bm
J^{\rm ph}.
\end{equation}
Therefore to obtain the change of angular momentum of electrons, one needs to subtract the contribution of phonon from the opposite change of lattice angular momentum. On the other hand one can measure the total magnetization change as
\begin{equation}\label{eq_deltm}
\Delta M=\Delta M^{\rm spin}+\Delta M^{\rm orb}.
\end{equation}
Combining Eq.~(\ref{eq_deltj}), Eq.~(\ref{eq_deltm}) together with the facts of
$\Delta M^{\rm orb}=\frac{e}{2m}\Delta J^{\rm orb}$ and $\Delta M^{\rm
spin}=\frac{e}{m}\Delta J^{\rm spin}$, one can easily determine $\Delta M^{\rm
spin}$ and $\Delta M^{\rm orbit}$.

The phonon can make a significant contribution to total angular momentum, while the magnitude of phonon angular momentum depends on the value of $\lambda$.  The parameter $\lambda$ can be obtained from phonon dispersion relation since our calculation shows that in the presence of spin-phonon interaction degenerate phonon modes split at $\Gamma$ point with a gap of $2\lambda$. By means of Raman scattering experiments, literatures \cite{split1,split2} show that the phonon splitting ranges up to about 26 cm$^{-1}$ in paramagnetic CeF$_3$ at $T=1.9$ K and $B=6$ T, thus $\lambda$ can be about $0.39$ THz and phonon angular momentum per unit cell is about 0.02 $ \hbar$.  One also can estimate the parameter $\lambda$ from phonon Hall effect. For a paramagnetic terbium gallium garnet Tb$_3$Ga$_5$O$_{12}$, the parameter $\lambda$ is estimated as $\lambda=0.1$ cm$^{-1}\simeq3$ GHz at $B=1$ T and $T=5.45$ K \cite{phet1}, thus in such material phonon angular momentum per unit cell is about $1.6\times10^{-4} \hbar$, which is relatively small. However, one can observe a much larger phonon angular momentum when magnetization is saturated in this paramagnetic material since the parameter $\lambda$ is proportional to magnetization.  In the phonon Hall effect experiment \cite{phee1,phee2}, the paramagnetic insulator was chosen to manifest the phonon contribution in the thermal transport where the contribution from electron and magnon can be neglected. However, the spin-phonon interaction is widely present in various magnetic materials \cite{spimat1, spimat2, spimat3,spimat4}. Ferromagnetic materials have very large magnetization, thus one can expect a large phonon angular momentum.   One also can observe evident phonon angular momentum in materials with strong spin-phonon interaction by using Raman spectroscopy,  such as La$_2$NiMnO$_6$ \cite{iliev07}, Sr$_2$CoO$_4$ \cite{pand13} and cupric oxide \cite{chen95}.

Thus for materials with strong spin-phonon interaction together with large magnetization, the zero-point angular momentum of phonons can be significant. According to previous studies, in some ferromagnetic materials the calculated orbital magnetic moment is only a few percent of the total magnetic moment, that is, the orbital angular momentum is also around a few percent of $\hbar$ \cite{reck69}, thus the phonon angular momentum can not be ignored. With improvement of experimental technique in past decades, accuracy of measurement has been much enhanced thus phonon angular momentum should be measurable.

{\it Possible experiment to separate phonon angular momentum} -- One can do
experiments on a ferromagnetic insulator with saturation magnetization, where
electron transport can be ignored. Due to the property of phonon angular momentum
-- it decreases with increasing temperature and vanishes in the classical limit,
one can measure the change of lattice angular momentum at low and high temperatures
to separate the phonon angular momentum from the others. Here the temperature scale
should be the Debye temperature which divides the quantum and classical regions.
On the other hand, in order to avoid the involvement of magnons, we need to
do experiments at temperatures low compared to the Curie temperature.  This demands
that the Curie temperature must be much higher than the Debye temperature.  Thus the angular momentum of magnons almost keeps constant while that of phonons changes
dramatically with changing temperature.  Fortunately, this can be satisfied by many
ferromagnetic materials where their Curie temperature are
around 1000 K while their Debye temperatures are less than 500 K \cite{kitt04}.

Besides application to measurement of gyromagnetic ratio,  nontrivial phonon angular momentum provides us a possible efficient route to study spin-phonon interaction in magnetic materials. On the other hand, to separate the contribution from phonons and magnons to thermal Hall effect in ferromagnetic materials is an open problem, phonon angular momentum can give a way to obtain the phonon contribution.

{\it Acknowledgements} -- We thank Junren Shi, Yang Gao, Zhenhua Qiao, Xiao Li, Ran Cheng for helpful discussions. We acknowledge support from  DOE-DMSE (DE-FG03-02ER45958),
NBRPC (2012CB-921300), NSFC (91121004), and the Welch Foundation (F-1255).

\newpage

\vskip 15 cm
\begin{widetext}
\setcounter{figure}{0}
\setcounter{equation}{0}
\setcounter{section}{0}

\renewcommand\thefigure{S\arabic{figure}}
\renewcommand\theequation{S\arabic{equation}}
\section{Supplementary information for ``Angular Momentum of Phonons and the Einstein-de Hass Effect''}

\vskip 0.5 cm

\subsection{Derivation of Eq.(2) and (3)}
The angular momentum of phonons is
\begin{eqnarray}
 J_z^{{\rm{ph}}} & = &\sum\limits_{l\alpha } {(u_{l\alpha }^x \dot u_{l\alpha }^y  - u_{l\alpha }^y \dot u_{l\alpha }^x )}  \nonumber \\
& = & \sum\limits_{l\alpha } {\left( {\begin{array}{*{20}c}
   {u_{l\alpha }^x }  \\
   {u_{l\alpha }^y }  \\
\end{array}} \right)^T \left( {\begin{array}{*{20}c}
   0 & 1  \\
   { - 1} & 0  \\
\end{array}} \right)\left( {\begin{array}{*{20}c}
   {\dot u_{l\alpha }^x }  \\
   {\dot u_{l\alpha }^y }  \\
\end{array}} \right).}  \\
\label{eq-s1}
\end{eqnarray}
For a unit cell with two atoms $n=2$, that is, $\alpha=1,2$, the angular momentum of phonons can be written as
{\small
\begin{equation}
J_z^{{\rm{ph}}} \; = \sum\limits_l {\left( {\begin{array}{*{20}c}
   {u_{l1}^x }  \\
   {u_{l1}^y }  \\
   {u_{l2}^x }  \\
   {u_{l2}^y }  \\
\end{array}} \right)^T\left( {\begin{array}{*{20}c}
   0 & 1 & {} & {}   \\
   { - 1} & 0 & {} & {}  \\
   {} & {} & 0 & 1  \\
   {} & {} & { - 1} & 0    \\
\end{array}} \right)} \left( {\begin{array}{*{20}c}
   {\dot u_{l1}^x }  \\
   {\dot u_{l1}^y }  \\
   {\dot u_{l2}^x }  \\
   {\dot u_{l2}^y }  \\
\end{array}} \right).
\end{equation}
}
Using the  $u_{l}=\left( \begin{smallmatrix} u_{l1}^x \; u_{l1}^y \; u_{l2}^x \;u_{l2}^y  \end{smallmatrix} \right)^T$,  we obtain
\begin{equation}
 J_z^{{\rm{ph}}}= \sum\limits_l u_l^T i M {\dot u_l}.
\end{equation}
where $M=\left( \begin{smallmatrix} 0 & -i \\ i & 0 \end{smallmatrix} \right)\otimes I_{n\times n} $.
By using the second quantization for $u_l$ as
\[u_l = \sum_{k} \epsilon_k e^{i ( {\bf R}_l \cdot {\bm k}-\omega_k t)} \sqrt{\frac{\hbar}{2\omega_k N}}\; a_k + {\rm h.c.},\]
we obtain
\begin{equation}
J_z^{{\rm{ph}}} =\frac{\hbar}{2N} \sum_l
\sum_{k,k'} \left( \sqrt{\frac{\omega_k}{\omega_{k'}}}\epsilon_k^\dagger M \epsilon_{k'} a_k^\dagger a_{k'}+
\sqrt{\frac{\omega_{k'}}{\omega_k}} \epsilon_{k'}^T (-M) \epsilon_{k}^* a_{k'} a_{k}^\dagger \right)
e^{i( \bm k'  - \bm k ){\bm R_l}} e^{i(\omega _k  - \omega _{k'} )t}.
\end{equation}

Here we ignore the $a\,a$ and $a^\dagger\,a^\dagger$ terms since they vary rapidly with time and have no contribution in equilibrium.

Since $\epsilon_{k'}^T (-M) \epsilon_{k}^*=\epsilon_k^\dagger M \epsilon_{k'}$ and $\frac{1}{N} \sum_l e^{i( \bm k'  - \bm k ){\bm R_l}}= \delta_{{\bm k},{\bm k}'}$, then
\begin{equation}
J_z^{{\rm{ph}}} =\frac{\hbar}{2}
\sum_{k,k'} \epsilon_k^\dagger M \epsilon_{k'}  \left( \sqrt{\frac{\omega_k}{\omega_{k'}}}a_k^\dagger a_{k'}+
\sqrt{\frac{\omega_{k'}}{\omega_k}}  a_{k'} a_{k}^\dagger \right)
\delta_{{\bm k},{\bm k}'} e^{i(\omega _k  - \omega _{k'} )t}.
\end{equation}
Due to the communication relation $[a_{\bm k,\sigma'}, a_{\bm k,\sigma}^\dagger]=\delta_{\sigma,\sigma'}$, we obtain Eq.(2) in the main text, that is
\begin{equation}
J_z^{{\rm{ph}}} =\frac{\hbar}{2}
\sum_{k,k'} \epsilon_k^\dagger M \epsilon_{k'}  \left( \sqrt{\frac{\omega_k}{\omega_{k'}}}+
\sqrt{\frac{\omega_{k'}}{\omega_k}}\right)a_k^\dagger a_{k'}
\delta_{{\bm k},{\bm k}'} e^{i(\omega _k  - \omega _{k'} )t} +\frac{\hbar}{2}\sum_k\epsilon_k^\dagger M \epsilon_{k} .
\end{equation}

In equilibrium we know $ \langle a_{\bm k,\sigma '}^{\dagger}a_{\bm k,\sigma }\rangle=f(\omega_{k})\delta_{\sigma,\sigma'} $, then we obtain
\begin{equation}
J_z^{{\rm{ph}}} =\hbar
\sum_{k} \epsilon_k^\dagger M \epsilon_{k} f(\omega_{k}) +\frac{\hbar}{2}\sum_k\epsilon_k^\dagger M \epsilon_{k},
\end{equation}
which is the Eq.(3) in the main text.

\subsection{Proof of the zero angular momentum of trivial phonon system without spin-phonon interaction}
For the trivial phonon system $H  = \frac{1}{2} p^Tp + \frac{1}{2} u^T K u$,
one has an eigenvalue problem as
\begin{equation}\label{eq-eps1}
 D(\bm k)\epsilon(\bm k, \sigma)=\omega_{\bm k, \sigma}^2 \epsilon(\bm k, \sigma). \end{equation}
Due to $D^\dagger=D$, then
\begin{equation}\label{eq-eps2}
  \epsilon^\dagger(\bm k, \sigma) D(\bm k)=\omega_{\bm k, \sigma}^2 \epsilon^\dagger(\bm k, \sigma).
\end{equation}
From Eq.\ref{eq-eps1}, for wave vector $- \bm k$, one has
\[ D(-\bm k)\epsilon(-\bm k, \sigma)=\omega_{-\bm k, \sigma}^2 \epsilon(-\bm k, \sigma), \]
then
\[ \epsilon^T(-\bm k, \sigma) D^T(-\bm k)=\omega_{-\bm k, \sigma}^2 \epsilon^T(-\bm k, \sigma). \]
And since $ D^T(-\bm k)=D^T(\bm k)$, then
\begin{equation}\label{eq-eps3}
  \epsilon^T(-\bm k, \sigma) D(\bm k)=\omega_{-\bm k, \sigma}^2 \epsilon^T(\bm k, \sigma).
\end{equation}
From Eq. (\ref{eq-eps2}) and Eq. (\ref{eq-eps3}), we can have
\begin{eqnarray}
  \omega_{-\bm k, \sigma} &=& \omega_{\bm k, \sigma} \\
  \epsilon(-\bm k, \sigma) &=& \epsilon^*(\bm k, \sigma) .
\end{eqnarray}
Therefore,
\begin{equation}
l_{-\bm k, \sigma}^z=(\epsilon^\dagger_{-\bm k, \sigma}\,M\,\epsilon_{-\bm k, \sigma})\hbar=(\epsilon^T_{\bm k, \sigma}\,M\,\epsilon^*_{\bm k, \sigma})\hbar=(\epsilon^\dagger_{\bm k, \sigma}\,M^T\,\epsilon_{\bm k, \sigma})\hbar,
\end{equation}
because of $M^T=-M$, then
\begin{equation}
 l_{-\bm k, \sigma}^z= -l_{\bm k, \sigma}^z.
\end{equation}
And $f(\omega_{-\bm k, \sigma})=f(\omega_{\bm k, \sigma})$, thus
\[J^{\rm ph}_z=\sum_{\sigma,\bm k} l_{\bm k, \sigma}^z\, [f(\omega_{\bm k, \sigma})+\frac{1}{2}]=0,\]
where the summation is over all the $\bm k$ points in the first Brillouin zone, and all the branches ($\omega_\sigma\geq 0$) are included.

\subsection{Another Proof of Eq. (3) for systems with a spin-phonon interaction }
In the presence of the Raman type spin phonon interaction, as stated in the main text, the Hamiltonian is
\begin{equation}\label{eq-ham}
H  = \frac{1}{2} (p-{ \tilde A }
u)^T (p-{ \tilde A }u) + \frac{1}{2} u^T K u.
\end{equation}
The polarization vector $\epsilon$ satisfies
\begin{equation}\label{eq-eigen}
\bigl[ (-i\omega + A)^2 + D\bigr] \epsilon = 0,
\end{equation}
where $D({\bf k}) = - A^2+\sum_{l'} K_{ll'} e^{i({\bf R}_{l'} -
{\bf R}_{l})\cdot {\bf k}}$ denotes the dynamic matrix and $A$ is
block diagonal with elements $\Lambda_\alpha$. Here, we use the short-hand notation $k=({\bf k},\sigma)$ to specify both the
wavevector and the phonon branch. $D, K_{l,l'},$ and $A$ are
all $nd\times nd$ matrices, where $n$ is the number of particles
in one unit cell and $d$ is the dimension of the vibration.

Equation (\ref{eq-eigen}) is not a standard eigenvalue problem.
However, we can describe the system by a polarization vector as
 $x =(\mu, \epsilon)^T$, where $\mu $ and $\epsilon $ are associated
  with the momenta and coordinates,
respectively. Using Bloch's theorem, one has
{\small $
i\frac{\partial }
{{\partial t}}x  = H_{\rm eff} x, \;
 H_{\rm eff}= i\left( \begin{smallmatrix} -A & -D \\
I_{nd } & -A \end{smallmatrix} \right),
$}
where the $I_{nd}$ is the $nd\times nd$ identity matrix.
Therefore, one obtains the eigenvalue problem of the equation of motion
{\small $
H_{\rm eff}\,x_k= \omega_k\,x_k, \;\;\;\;
\tilde{x}_k^T\,H_{\rm eff} = \omega_k\,\tilde{x}_k^T,
$}
where $x_k =(\mu_k,
\epsilon_k)^T$ is the right eigenvector, and the left eigenvector is chosen as ${\tilde x}_k^T = (\epsilon^\dagger_k, - \mu^\dagger_k)/( - 2i\omega
_k)$, in such choice the second quantization of the Hamiltonian is satisfied.  The
orthonormal condition holds between the left and right
eigenvectors, as $ {\tilde
x_{\sigma,{\bf k}}}^T\;x_{\sigma',{\bf k}}=\delta_{\sigma\sigma'}$. We also have the completeness relation as $\sum_\sigma x_{\sigma,{\bf k}}\otimes{\tilde
x_{\sigma,{\bf k}}}^T= I_{2nd }$. The normalization of the eigenmodes is equivalent to  $\epsilon_k^\dagger\,\epsilon_k + \frac{i}{\omega_k} \epsilon_k^\dagger\, A\, \epsilon_k  = 1$.

From the eigenvalue problem Eq.~(\ref{eq-eigen}), we know that
the completed set contains branches of negative frequency.
The short-hand notation $k=({\bf k},\sigma)$ can include negative branches, $-k$ means $(-{\bf k},-\sigma)$. In order to simplify the notation, for all the branches, we define $a_{-k}=a_{k}^{\dagger}$.
The time dependence of the operators is given by:
$
a_{k}(t)=a_{k} e^{-i \omega_{k}t},
a_{k}^{\dagger}(t)= a_{k}^{\dagger} e^{i \omega_{k}t}.
$

Inserting the equation of motion $ \dot u_l = p_l - A u_l$, that is
\begin{equation}
\left( {\begin{array}{*{20}c}
   {\dot u_{l\alpha }^x }  \\
   {\dot u_{l\alpha }^y }  \\
\end{array}} \right) = \left( {\begin{array}{*{20}c}
   {p_{l\alpha }^x }  \\
   {p_{l\alpha }^y }  \\
\end{array}} \right) - \left( {\begin{array}{*{20}c}
   0 & \lambda_{l\alpha }  \\
   { - \lambda_{l\alpha } } & 0  \\
\end{array}} \right)\left( {\begin{array}{*{20}c}
   {u_{l\alpha }^x }  \\
   {u_{l\alpha }^y }  \\
\end{array}} \right),
\end{equation}
Eq.~\ref{eq-s1} becomes
\begin{eqnarray}
 J_z^{{\rm{ph}}} & = & \sum\limits_{l\alpha } {\left( {\begin{array}{*{20}c}
   {u_{l\alpha }^x }  \\
   {u_{l\alpha }^y }  \\
\end{array}} \right)^T \left( {\begin{array}{*{20}c}
   0 & 1  \\
   { - 1} & 0  \\
\end{array}} \right)} \left( {\begin{array}{*{20}c}
   {p_{l\alpha }^x }  \\
   {p_{l\alpha }^y }  \\
\end{array}} \right) + \left( {\begin{array}{*{20}c}
   {u_{l\alpha }^x }  \\
   {u_{l\alpha }^y }  \\
\end{array}} \right)^T \left( {\begin{array}{*{20}c}
   \lambda_{l\alpha }  & 0  \\
   0 & \lambda_{l\alpha }   \\
\end{array}} \right)\left( {\begin{array}{*{20}c}
   {u_{l\alpha }^x }  \\
   {u_{l\alpha }^y }  \\
\end{array}} \right) \nonumber \\
& = & \sum\limits_{l\alpha } {\left( {\begin{array}{*{20}c}
   {u_{l\alpha }^x }  \\
   {u_{l\alpha }^y }  \\
   { - p_{l\alpha }^x }  \\
   { - p_{l\alpha }^y }  \\
\end{array}} \right)^T} \left( {\begin{array}{*{20}c}
   0 & 1 & \lambda_{l\alpha }  & 0  \\
   { - 1} & 0 & 0 & \lambda_{l\alpha }   \\
   0 & 0 & 0 & 0  \\
   0 & 0 & 0 & 0  \\
\end{array}} \right)\left( {\begin{array}{*{20}c}
   {p_{l\alpha }^x }  \\
   {p_{l\alpha }^y }  \\
   {u_{l\alpha }^x }  \\
   {u_{l\alpha }^y }  \\
\end{array}} \right). \\
 \end{eqnarray}
For a unit cell with two atoms $n=2$, the angular momentum of phonons can be written as
{\small
\begin{equation}
J_z^{{\rm{ph}}} \; = \sum\limits_l {\left( {\begin{array}{*{20}c}
   {u_{l1}^x }  \\
   {u_{l1}^y }  \\
   {u_{l2}^x }  \\
   {u_{l2}^y }  \\
   { - p_{l1}^x }  \\
   { - p_{l1}^y }  \\
   { - p_{l2}^x }  \\
   { - p_{l2}^y }  \\
\end{array}} \right)^T\left( {\begin{array}{*{20}c}
   0 & 1 & {} & {} & \lambda_{l1 }  & 0 & {} & {}  \\
   { - 1} & 0 & {} & {} & 0 & \lambda_{l1 }  & {} & {}  \\
   {} & {} & 0 & 1 & {} & {} & \lambda_{l2 }  & 0  \\
   {} & {} & { - 1} & 0 & {} & {} & 0 & \lambda_{l2 }   \\
   {} & {} & {} & {} & {} & {} & {} & {}  \\
   {} & {} & {} & {} & {} & {} & {} & {}  \\
   {} & {} & {} & {} & {} & {} & {} & {}  \\
   {} & {} & {} & {} & {} & {} & {} & {}  \\
\end{array}} \right)} \left( {\begin{array}{*{20}c}
   {p_{l1}^x }  \\
   {p_{l1}^y }  \\
   {p_{l2}^x }  \\
   {p_{l2}^y }  \\
   {u_{l1}^x }  \\
   {u_{l1}^y }  \\
   {u_{l2}^x }  \\
   {u_{l2}^y }  \\
\end{array}} \right).
\end{equation}
}
Using the  $\chi_{l}=\left( \begin{smallmatrix} p_{l} \\ u_{l} \end{smallmatrix} \right)$ and $\tilde{\chi}_{l} = \left( \begin{smallmatrix}u_{l} \\ -p_{l} \end{smallmatrix} \right)$, where $u_l, p_l$ are column vectors of displacements and conjugate momenta for the $l-$th unit, if $n=2$, then $
u_l  = \left( {\begin{array}{*{20}c}
   {u_{l1}^x } & {u_{l1}^y } & {u_{l2}^x } & {u_{l2}^y }  \\
\end{array}} \right)^T $, similar for $p_l$. Then we obtain  Eq. (3) in the main text as
\begin{equation}
J^{\rm ph}_z=\sum_{l} \tilde{\chi}^T_{l} \left( \begin{array}{cc} iM & -iMA \\
0 & 0 \end{array} \right) \chi_{l}.
\end{equation}

By the second quantization $ \chi_{l}=\sqrt{\frac{\hbar}{N}}\sum_k x_k e^{i {\bf R}_l \cdot {\bf k}} \sqrt{\frac{1}{2 |\omega_{k}|}} \; a_k; \;\; \tilde{\chi}_{l} = \sqrt{\frac{\hbar}{N}}\sum_k \tilde x_k e^{-i {\bf R}_l \cdot {\bf k}} ( - 2i\omega _k )\sqrt{\frac{1}{2|\omega_{k}| }}\; a_k^{\dagger}$, and $ x_k =  \left( \begin{smallmatrix} \mu_{k} \\ \epsilon_k\end{smallmatrix} \right)$, $k=({\bf k},\sigma)$, $\tilde x_{k}^T = \frac{1}{- 2i\omega_k}\left( \begin{smallmatrix} \epsilon^\dagger_{k}\;&- \mu^\dagger_k \end{smallmatrix} \right)$, we obtain
\begin{equation}
J_z^{{\rm{ph}}} \; = \sum\limits_{l,k,k'} {e^{i{\bm R}_l (\bm k - \bm k')} } \frac{\hbar }{{ N}}\frac{1}{{2\sqrt {|\omega _{k'} ||\omega _k |} }}\left( {\begin{array}{*{20}c}
   {\varepsilon _{k'}^\dag  } & { - \mu _{k'}^\dag  }  \\
\end{array}} \right)\left( {\begin{array}{*{20}c}
   iM & { -iMA }  \\
   0 & 0  \\
\end{array}} \right)\left( {\begin{array}{*{20}c}
   {\mu _k }  \\
   {\varepsilon _k }  \\
\end{array}} \right)a_{k'}^\dag  a_k.
\end{equation}
By using the fact of $ \sum\limits_l {e^{i{\bm R}_l (\bm k - \bm k')} }  = N\delta _{\bm k,\bm k'} $, we obtain

\begin{equation}
J_z^{{\rm{ph}}} \; = \sum\limits_{\bm k,\sigma ,\sigma '} {\frac{\hbar }{{2\sqrt {|\omega _{\bm k,\sigma } ||\omega _{\bm k,\sigma '} |} }}\left( {\begin{array}{*{20}c}
   {\varepsilon _{\bm k,\sigma '}^\dag  } & { - \mu _{\bm k,\sigma '}^\dag  }  \\
\end{array}} \right)\left( {\begin{array}{*{20}c}
   iM & { - iMA }  \\
   0 & 0  \\
\end{array}} \right)\left( {\begin{array}{*{20}c}
   {\mu _{\bm k,\sigma } }  \\
   {\varepsilon _{\bm k,\sigma } }  \\
\end{array}} \right)a_{\bm k,\sigma '}^\dag  a_{\bm k,\sigma } }.
\end{equation}
In equilibrium, we know that {\small $ \langle a_{\bm k,\sigma '}^{\dagger}a_{\bm k,\sigma }\rangle=f(\omega_{k}){\rm sign}(\sigma)\delta_{\sigma,\sigma'} $}, then
\begin{equation}
\langle J_z^{{\rm{ph}}} \rangle \; = \sum\limits_{k} {\frac{\hbar }{{2|\omega _{k } |} }}\left( {\begin{array}{*{20}c}
   {\varepsilon _{k}^\dag  } & { - \mu _{k}^\dag  }  \\
\end{array}} \right)\left( {\begin{array}{*{20}c}
   iM & { - iMA }  \\
   0 & 0  \\
\end{array}} \right)\left( {\begin{array}{*{20}c}
   {\mu _{k } }  \\
   {\varepsilon _{k } }  \\
\end{array}} \right) \langle a_{k}^\dag  a_{k }\rangle .
\end{equation}

And using the relation between the momentum and displacement polarization vectors $ \mu_k = -i\omega_k \epsilon_k + A\epsilon_k$, we get

\begin{equation}
\left( {\begin{array}{*{20}c}
   {\varepsilon _k^\dag  } & { - \mu _k^\dag  }  \\
\end{array}} \right)\left( {\begin{array}{*{20}c}
   iM & { - iMA }  \\
   0 & 0  \\
\end{array}} \right)\left( {\begin{array}{*{20}c}
   {\mu _k }  \\
   {\varepsilon _k }  \\
\end{array}} \right) = \omega \varepsilon _k^\dag  M\varepsilon _k.
\end{equation}
Then the phonon angular momentum in equilibrium is
\begin{equation}
J^{\rm ph}_z=\frac{ \hbar}{2 }  \sum_{k}  \epsilon^\dagger_{k}\,M\,\epsilon_{k}\, f(\omega_k),
\end{equation}
by using $\omega_k{\rm sign}(\sigma)=|\omega_k|$.
 \begin{figure}[t]
\includegraphics[width=5 in,  angle=0]{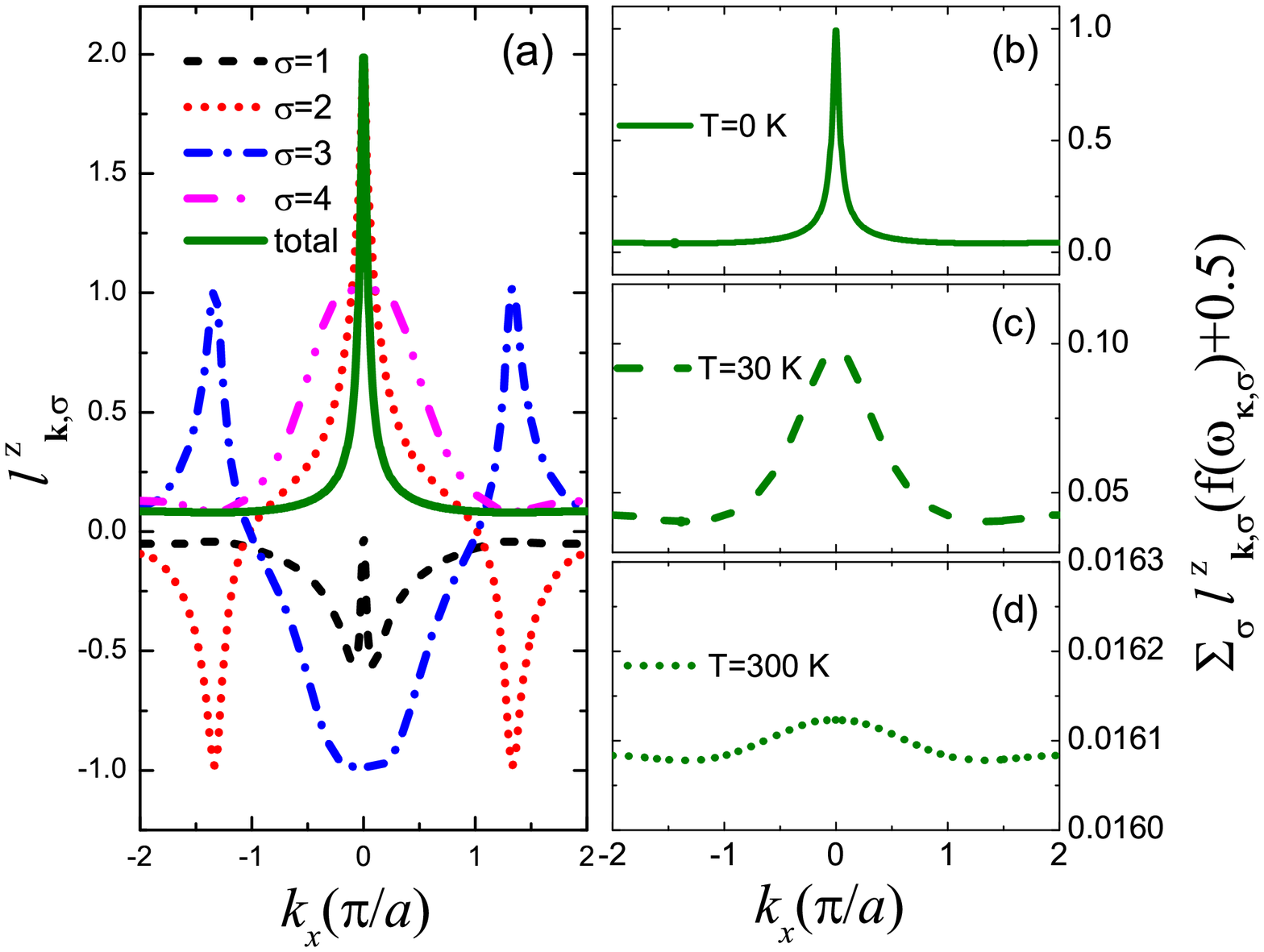}
\caption{ \label{figS1} (Color online) (a) The angular momentum $l_{\bm k, \sigma}^z$ of bands $\sigma=1$ - 4 as a function of $k_x$. (b)-(d) The sum of the angular momentum of four bands $\sum_{\sigma>0} l_{\bm k, \sigma}^z\, [f(\omega_{\bm k, \sigma})+\frac{1}{2}]$ at different temperatures.   The parameters are the same with those in the main text. }
\end{figure}
We have $\epsilon_{-\bm k, -\sigma}=\epsilon^*_{\bm k, \sigma}$, then
\begin{equation}
(\epsilon^\dagger_{-\bm k, -\sigma}\,M\,\epsilon_{-\bm k, -\sigma})\hbar=(\epsilon^T_{\bm k, \sigma}\,M\,\epsilon^*_{\bm k, \sigma})\hbar=(\epsilon^\dagger_{\bm k, \sigma}\,M^T\,\epsilon_{\bm k, \sigma})\hbar=-(\epsilon^\dagger_{\bm k, \sigma}\,M\,\epsilon_{\bm k, \sigma})\hbar,
\end{equation}
but $f(\omega_{-\bm k, -\sigma})\neq f(\omega_{\bm k, \sigma})$, thus phonon angular momentum at $-\bm k, -\sigma$ cannot cancel out that at $\bm k, \sigma$, and we can get a nonzero phonon angular momentum.
Here, $k=({\bm k},\sigma)$ includes all the positive $\sigma>0$ and negative $\sigma<0$ branches.  For the negative branches, that is $\sigma<0, \omega<0$,   using {\small $\epsilon_{-k}^* = \epsilon_{k}; \;\omega_{-k} = -\omega_{k}$ }, then
\[
\frac{\hbar }{2}\sum\limits_{\bm k,\sigma  < 0} {\varepsilon _{\bm k,\sigma }^\dag  M\varepsilon _{\bm k,\sigma } f(\omega _{\bm k,\sigma } )}  = \frac{\hbar }{2}\sum\limits_{\bm k',\sigma ' > 0} {\varepsilon _{ - \bm k', - \sigma '}^\dag  M\varepsilon _{ - \bm k', - \sigma '} f(\omega _{ - \bm k', - \sigma '} )}  = \frac{\hbar }{2}\sum\limits_{\bm k',\sigma ' > 0} {\varepsilon _{\bm k',\sigma '}^T M\varepsilon _{\bm k',\sigma '}^* f( - \omega _{\bm k',\sigma '} )}.
\]
And $\varepsilon _{\bm k',\sigma '}^T M\varepsilon _{\bm k',\sigma '}^*=\varepsilon _{\bm k',\sigma '}^\dagger M^T\varepsilon _{\bm k',\sigma '}$, $M^T=-M$, $f( - \omega _{\bm k',\sigma '} =-(1+f(  \omega _{\bm k',\sigma '} )$, then
\[
\frac{\hbar }{2}\sum\limits_{\bm k,\sigma  < 0} {\varepsilon _{\bm k,\sigma }^\dag  M\varepsilon _{\bm k,\sigma } f(\omega _{\bm k,\sigma } )}  = \frac{\hbar }{2}\sum\limits_{\bm k',\sigma ' > 0} {\varepsilon _{\bm k',\sigma '}^\dag  M\varepsilon _{\bm k',\sigma '}^{} (f(\omega _{\bm k',\sigma '} )}  + 1).
\]
Therefore we obtain the angular momentum of phonons as:
\begin{equation}\label{eq_angph}
J^{\rm ph}_z=\sum_{\sigma>0,\bm k} l_{\bm k, \sigma}^z\, [f(\omega_{\bm k, \sigma})+\frac{1}{2}],\;\; l_{\bm k, \sigma}^z=(\epsilon^\dagger_{\bm k, \sigma}\,M\,\epsilon_{\bm k, \sigma})\hbar.
\end{equation}
Here due to $\epsilon_{-\bm k, \sigma}=\epsilon_{\bm k, -\sigma}^* \neq \epsilon_{\bm k, \sigma}^*$, we cannot obtain $ l_{-\bm k, \sigma}^z= -l_{\bm k, \sigma}^z$, thus the phonon angular momentum at the wave vector $-\bm k$ cannot cancel out that at $\bm k$, and we can obtain a nonzero angular momentum of phonons.

We show the phonon angular momentum as a function of $k_x$ ($k_y=0$) in Fig. \ref{figS1}. Due to the 6-fold symmetry of the honeycomb lattice, the angular momentum of phonons is an even function of $k_x$, as shown in Fig. \ref{figS1}. For band 1 the angular momenta of all the wave-vector points are negative, while they are positive for band 4; for both band 2 and 3, the phonon angular momentum (by doing the summation over all the four bands) at $\bm \Gamma$ point is opposite to that at $\bm K$ point. The total angular momentum of four branches has a maximum at $\bm \Gamma$ point, and it decreases as $k_x$ increasing. The phonon angular momentum arrives at its minimum at the point $\bm K$ ($k=4\pi/3$) with a positive value around $0.08 \hbar$. The phonons at every $\bm k$ points in the Brillouin zone has nonzero phonon angular momentum.  The difference of the total phonon angular momentum between those at $\bm \Gamma$  and $\bm K$ will be smoothed by the temperature, which can be seen in  Fig. \ref{figS1} (b)(c)(d).  When $T=0$ K, the phonon angular momentum $\bm \Gamma$ is around 20 times of that at $\bm K$ (Fig. \ref{figS1} (b)), and the ratio decreases to about 2 when $T=30K$ (see  Fig. \ref{figS1} (c)). When $T=300$ K and the phonon angular momentum will be almost the same for all the wave-vector points, as shown in Fig. \ref{figS1} (d).

%We plot the phonon angular momentum for different band as function of temperature in Fig. \ref{figS2}(a) and (b). At low temperatures, the phonon angular momentum from the acoustic bands (band 1 and 2) almost cancel out each other, thus the total phonon angular momentum mostly comes from the contribution of optical bands (band 3 and 4). Form Fig. \ref{figS2}(c) and (d), we can see that the sum of phonon angular momenta of the acoustic bands almost vanishes if $\lambda$ is not too large, when the optical bands dominate the contribution to total phonon angular momentum.
%\begin{figure}[b]
%\includegraphics[width=5 in,  angle=0]{figS2.eps}
%\caption{ \label{figS2} (Color online) The phonon angular momentum $J_z^{ph}$ of one unit cell from different phonon band as a function of temperature
%$T$ ((a) and (b)) and of $\lambda$ ((c) and (d)). For (a) and (b), $\lambda=$ 1 THz; for (c) and (d), $T=0$.    Other parameters are the same with those in the main text. }
%\end{figure}

\subsection{ phonon angular momentum as a odd function of $\lambda$}

\begin{eqnarray}
 J_z^{{\rm{ph}}} ( - \lambda ) & = &\hbar \sum\limits_{\bm k,\sigma  > 0} {\varepsilon _{\bm k,\sigma }^\dag  ( - \lambda )M\varepsilon _{\bm k,\sigma }^{} ( - \lambda )(f(\omega _{\bm k,\sigma } ( - \lambda ))}  + 1) \nonumber \\
& \Downarrow &  \varepsilon _{ - \bm k, \sigma}^* ( - \lambda ) = \varepsilon _{\bm k, \sigma} (\lambda )\quad \omega _{ - \bm k, \sigma}^{} ( - \lambda ) = \omega _{\bm k, \sigma} (\lambda ) \nonumber \\
& =& \hbar \sum\limits_{\bm k,\sigma  > 0} {\varepsilon _{ - \bm k,\sigma }^T (\lambda )M\varepsilon _{ - \bm k,\sigma }^* (\lambda )(f(\omega _{ - \bm k,\sigma } (\lambda ))}  + 1) \nonumber \\
 & \Downarrow & \varepsilon _{ - \bm k,\sigma }^T (\lambda )M\varepsilon _{ - \bm k,\sigma }^* (\lambda ) = \varepsilon _{ - \bm k,\sigma }^\dag  (\lambda )M^T \varepsilon _{ - \bm k,\sigma }^{} (\lambda ) =  - \varepsilon _{ - \bm k,\sigma }^\dag  (\lambda )M\varepsilon _{ - \bm k,\sigma }^{} (\lambda ) \nonumber \\
& =&  - \hbar \sum\limits_{\bm k',\sigma  > 0} {\varepsilon _{\bm k',\sigma }^\dag  (\lambda )M\varepsilon _{\bm k',\sigma }^{} (\lambda )(f(\omega _{\bm k',\sigma } (\lambda ))}  + 1) \nonumber \\
& =&  - J_z^{{\rm{ph}}} (\lambda ).
  \end{eqnarray}

\subsection{ phonon angular momentum in the classical limit}

At a high temperature, we can expand the Einstein-Bose distribution in Taylor series as

\begin{equation}
f(x) = \frac{1}{{e^x  - 1}} \simeq \frac{1}{x} - \frac{1}{2} + \frac{1}{{12}}x + O(x^2 ),\quad x = \frac{{\hbar \omega }}{{k_B T}}.
\end{equation}
Inserting it to Eq. (5) in the main text, we have
\begin{equation}
J^{\rm ph}_z(T\rightarrow\infty)=\sum_{\sigma>0,\bm k} [ (\frac{k_B T}{\hbar\omega}+\frac{\hbar\omega}{12 k_B T} )l_{\bm k, \sigma}^z ],
\end{equation}
The term linear to $T$ is zero, that is, $ \sum_{\sigma>0,\bm k} \frac{\epsilon^\dagger_{\bm k, \sigma}\,M\,\epsilon_{\bm k, \sigma}}{\omega_{\bm k,\sigma}}=0$,   we prove it as follows.

Since we have the completeness relation as
\begin{equation}
 \sum_\sigma x_{\bm k,\sigma}\otimes{\tilde
x_{\bm k,\sigma}}^T= I_{4n\times4n},
\end{equation}
with $ x_k =  \left( \begin{smallmatrix} \mu_{k} \\ \epsilon_k\end{smallmatrix} \right)$, $\tilde x_{k}^T = \frac{1}{- 2i\omega_k}\left( \begin{smallmatrix} \epsilon^\dagger_{k}\;&- \mu^\dagger_k \end{smallmatrix} \right)$.
Thus, the off-diagonal block
\begin{equation}
 \sum_\sigma \epsilon_{\bm k,\sigma}\otimes{\tilde
\epsilon_{\bm k,\sigma}}^\dagger/(-2i\omega_{\bm k,\sigma})= O_{2n\times2n},
\end{equation}
where $O_{2n\times2n}$ is a zero matrix. Then we have, for arbitrary $i$ and $j$,
\begin{equation}
 \sum_\sigma \frac{\epsilon_i(\bm k,\sigma)\epsilon_j^*(\bm k,\sigma)}{\omega_{\bm k,\sigma}}=0.
\end{equation}
Similar as that in Section II, we can prove
\[
\sum\limits_{\bm k,\sigma  < 0} {\frac{{\varepsilon _{\bm k,\sigma }^\dag  M\varepsilon _{\bm k,\sigma }^{} }}{{\omega _{\bm k,\sigma }^{} }}}  = \sum\limits_{\bm k',\sigma ' > 0} {\frac{{\varepsilon _{ -\bm  k', - \sigma '}^\dag  M\varepsilon _{ - \bm k', - \sigma '}^{} }}{{\omega _{\bm k', - \sigma '}^{} }}}  = \sum\limits_{\bm k',\sigma ' > 0} \frac{{\varepsilon _{\bm k',\sigma '}^T M\varepsilon _{\bm k',\sigma '}^* }}{{ - \omega _{\bm k',\sigma '}^{} }} = \sum\limits_{\bm k,\sigma  > 0} {\frac{{\varepsilon _{\bm k,\sigma }^\dag  M\varepsilon _{\bm k,\sigma }^{} }}{{\omega _{\bm k,\sigma }^{} }}}.
\]
Then
\begin{equation}
\sum\limits_{\bm k,\sigma  > 0} {\frac{{\varepsilon _{\bm k,\sigma }^\dag  M\varepsilon _{\bm k,\sigma }^{} }}{{\omega _{\bm k,\sigma }^{} }}}  = \frac{1}{2}\sum\limits_{\bm k,\sigma } {\frac{{\varepsilon _{\bm k,\sigma }^\dag  M\varepsilon _{\bm k,\sigma }^{} }}{{\omega _{\bm k,\sigma }^{} }}}  = \frac{1}{2}\sum\limits_{\scriptstyle \bm k,\sigma  \hfill \atop
  \scriptstyle j,i \hfill} {\frac{{\varepsilon _j^* (\bm k,\sigma )M_{ji} \varepsilon _i^{} (\bm k,\sigma )}}{{\omega _{\bm k,\sigma }^{} }}}  = \frac{1}{2}\sum\limits_{\scriptstyle \bm k \hfill \atop
  \scriptstyle j,i \hfill} {M_{ji} \sum\limits_\sigma  {\frac{{\varepsilon _j^* (\bm k,\sigma )\varepsilon _i^{} (\bm k,\sigma )}}{{\omega _{\bm k,\sigma }^{} }}}  = } 0.
\end{equation}

\end{widetext}


\begin{thebibliography}{01}
\bibitem{comment2} If $\lambda_\alpha$ is different for different site, one still can get nonero phonon angular momentum. Due to the rescaling for coordinate and momentum in the Hamiltonian by ionic mass $m_\alpha$, $\lambda_\alpha$ is proportional to $1/m_\alpha$. If one unit cell has two sites and $m_1\lambda_1=-m_2\lambda_2$, the phonon angular momentum is nonzero for $m_1\neq m_2$ and zero for $m_1=m_2$ when the contribution form the two sites exactly cancels out.
\end{thebibliography}
\end{document}